\documentclass[sigconf,nonacm]{acmart}

\setcopyright{none}
\renewcommand\footnotetextcopyrightpermission[1]{}
\acmConference[KDD '26 Workshop]{KDD 2026 Workshop on
  Agentic Software Engineering}{August 10, 2026}{Jeju, Korea}

\usepackage{xcolor}
\usepackage{booktabs}
\usepackage{float}
\usepackage{xspace}
\usepackage{graphicx}
\usepackage{microtype}
\usepackage{tikz}
\usepackage{adjustbox}
\usepackage{tabularx}
\usetikzlibrary{arrows.meta, positioning, fit, backgrounds}
\newcommand{\atm}{\textsc{Atm}\xspace}
\newcommand{\full}{\checkmark}
\newcommand{\partl}{$\triangle$}
\newcommand{\none}{$\times$}

\begin{document}

\title{Tracing the Messy Middle: \\
       A Runtime Transaction Schema for AI Coding Agents}

\author{Yuelin Hou}
\authornote{Work done while interning at QingCheng.AI.}
\affiliation{%
  \institution{QingCheng.AI}
  \city{Shanghai}
  \country{China}
}
\affiliation{%
  \institution{Duke University}
  \city{Durham}
  \country{United States}
}

\author{Haorui Li}
\authornote{Corresponding author.}
\affiliation{%
  \institution{QingCheng.AI}
  \city{Shanghai}
  \country{China}
}
\email{lihaorui@qingcheng.ai}

\author{Lu Chen}
\authornotemark[1]
\affiliation{%
  \institution{QingCheng.AI}
  \city{Shanghai}
  \country{China}
}
\affiliation{%
  \institution{Fudan University}
  \city{Shanghai}
  \country{China}
}

\author{Zhen Chen}
\authornotemark[1]
\affiliation{%
  \institution{QingCheng.AI}
  \city{Shanghai}
  \country{China}
}
\affiliation{%
  \institution{University of Science and Technology of China}
  \city{Hefei}
  \country{China}
}

\begin{abstract}
AI coding agents are increasingly integrated into software development, yet the field lacks a systematic feedback loop to improve their behavior in practice. Artifact datasets capture final outputs such as code patches, commits, and benchmark outcomes, but omit the messy middle of the development lifecycle: the failed retries, fallback strategies, routing decisions, and abandoned sessions that determine operational cost, reliability, and productivity.

We propose an observability framework that captures these runtime behaviors as structured transaction records, enabling engineering coaching grounded in behavioral evidence, workflow optimization at the stages where tasks break down, and routing policy iteration measured by total task cost rather than per-request price. We define the runtime transaction schema this architecture requires and demonstrate initial feasibility through Agent Token Manager (\atm), a local-first prototype that reconciles gateway records and transcript evidence without exposing prompt content to external collectors. Finally, we identify three open research problems that must be solved to scale these feedback mechanisms: trace reconciliation without shared request identifiers, outcome linking across asynchronous systems, and intervention causality. Treating the linkage between sessions and artifacts as a fundamental infrastructure requirement provides a structured foundation for understanding and optimizing agent behavior.
\end{abstract}


\maketitle

\section{Introduction}
\label{sec:intro}

Consider three scenarios that recur in agentic development practice.

\textbf{Vignette 1.} A developer's Claude Code session consumes
substantial cost and produces no commit; CI, git, and PR queues are
empty, so by every artifact-level signal nothing happened. Yet at
runtime the agent hit a rate limit, emitted a malformed tool call,
fell back to a cheaper model that returned an incorrect patch, and
terminated after several failed repairs, invisible to any artifact
dataset.

\textbf{Vignette 2.} A team deploys a cost-aware routing policy:
tasks above 8K tokens route to a cheaper provider, and per-request
price drops 60\%. Two weeks later, total agent spend has increased.
The gateway records only the resolved model and token count; it
omits retry chains, fallback frequency, and whether each task
produced an accepted patch~\cite{frugalgpt2023, routellm2024}.

\textbf{Vignette 3.} A senior and a junior engineer use the same
agent on the same class of bug: the senior lands accepted patches in
two to three requests; the junior averages nine with three aborts.
The difference is visible in runtime traces: broad context
selection, absent plan steps, and retries without new information.
Yet no system connects this signal to a coaching recommendation.

Each vignette describes a situation where the information needed to
diagnose and fix the problem existed at runtime but was never
captured in a form that supports action. This is not a shortage of
analytics methods. Funnel analysis, cohort comparison, and policy
A/B testing are mature techniques~\cite{tang2010overlapping,
kohavi2013online}; the missing piece is the \emph{data substrate},
the runtime transaction layer that would make these techniques
applicable to AI teammates. Without that layer, teams are left
choosing between ignoring the problem and guessing at the fix.

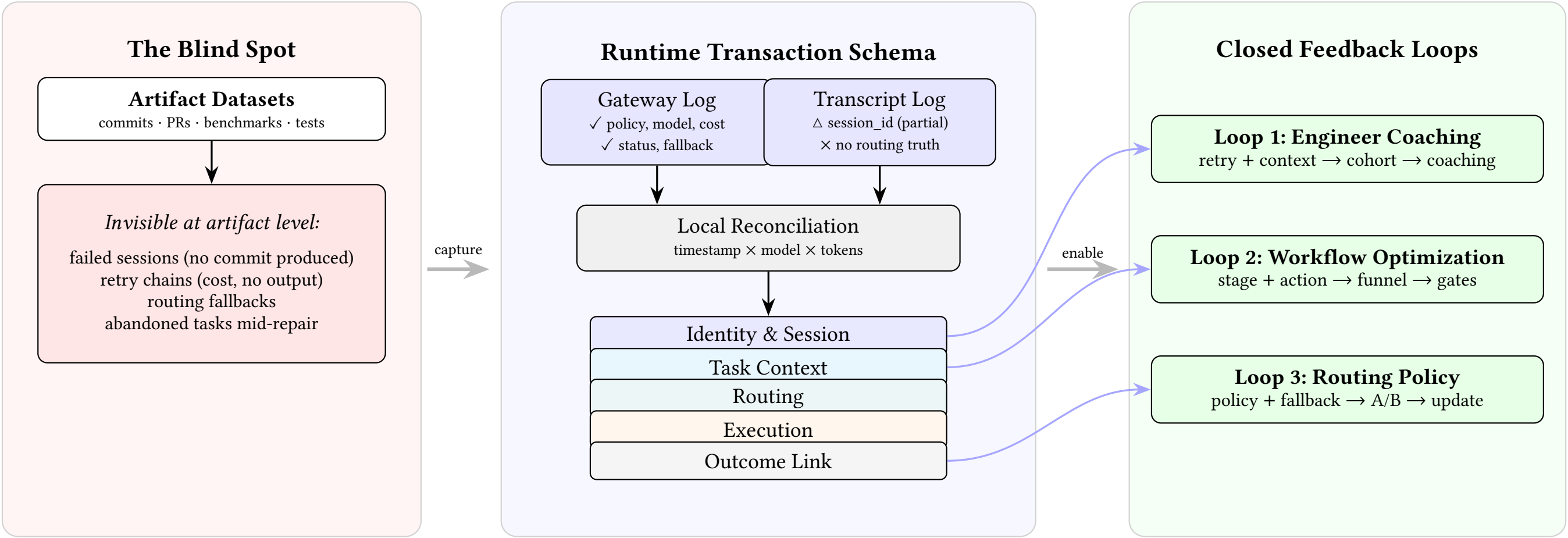
\begin{figure*}[!t]
\centering
\begin{adjustbox}{width=\textwidth}
\begin{tikzpicture}[
  font=\footnotesize,
  arr/.style={-Stealth, semithick},
  bigarr/.style={-Stealth, very thick, draw=gray!55},
  panelbg/.style={draw=gray!35, rounded corners=5pt, inner sep=7pt},
  srcbox/.style={draw, rounded corners=3pt, align=center, inner sep=4pt,
                 minimum width=2.6cm, minimum height=0.65cm},
  layerbox/.style={draw, rounded corners=2pt, align=center, inner sep=3pt,
                   minimum width=4.0cm, minimum height=0.42cm},
  loopbox/.style={draw, rounded corners=3pt, align=center, inner sep=4pt,
                  minimum width=4.4cm, minimum height=0.72cm},
]

\node[panelbg, fill=red!4, minimum width=4.7cm, minimum height=6.0cm]
  (p1bg) at (0,0) {};
\node[font=\small\bfseries] at (0, 2.45) {The Blind Spot};

\node[srcbox, fill=white, minimum width=3.9cm] at (0, 1.8) (artbox) {
  \textbf{Artifact Datasets}\\[-1pt]
  {\tiny commits~$\cdot$~PRs~$\cdot$~benchmarks~$\cdot$~tests}
};

\node[srcbox, fill=red!10, minimum width=3.9cm, minimum height=2.0cm]
  at (0, -0.05) (missbox) {
  \textit{Invisible at artifact level:}\\[3pt]
  {\scriptsize failed sessions (no commit produced)}\\[-1pt]
  {\scriptsize retry chains (cost, no output)}\\[-1pt]
  {\scriptsize routing fallbacks}\\[-1pt]
  {\scriptsize abandoned tasks mid-repair}
};

\draw[arr] (artbox.south) -- (missbox.north);

\draw[bigarr] (2.42,0) -- (3.12,0)
  node[midway, above, font=\tiny] {capture};

\node[panelbg, fill=blue!3, minimum width=6.0cm, minimum height=6.0cm]
  (p2bg) at (6.25,0) {};
\node[font=\small\bfseries] at (6.25, 2.45) {Runtime Transaction Schema};

\node[srcbox, fill=blue!10] at (5.0, 1.65) (gw) {
  Gateway Log\\[-1pt]{\tiny \checkmark~policy, model, cost}\\[-1pt]{\tiny \checkmark~status, fallback}
};
\node[srcbox, fill=blue!10] at (7.5, 1.65) (tr) {
  Transcript Log\\[-1pt]{\tiny $\triangle$~session\_id (partial)}\\[-1pt]{\tiny $\times$~no routing truth}
};

\node[srcbox, fill=gray!12, minimum width=4.3cm] at (6.25, 0.35) (rec) {
  Local Reconciliation\\[-1pt]{\tiny timestamp $\times$ model $\times$ tokens}
};
\draw[arr] (gw.south) -- (gw.south |- rec.north);
\draw[arr] (tr.south) -- (tr.south |- rec.north);

\node[layerbox, fill=blue!9]   at (6.25, -0.75) (lid)  {Identity~\&~Session};
\node[layerbox, fill=cyan!8]   at (6.25, -1.10) (ltask){Task Context};
\node[layerbox, fill=teal!7]   at (6.25, -1.45) (lrout){Routing};
\node[layerbox, fill=orange!7] at (6.25, -1.80) (lexec){Execution};
\node[layerbox, fill=gray!8]   at (6.25, -2.15) (lout) {Outcome Link};
\draw[arr] (rec.south) -- (lid.north);

\draw[bigarr] (9.38,0) -- (10.18,0)
  node[midway, above, font=\tiny] {enable};

\node[panelbg, fill=green!4, minimum width=4.9cm, minimum height=6.0cm]
  (p3bg) at (12.75,0) {};
\node[font=\small\bfseries] at (12.75, 2.45) {Closed Feedback Loops};

\node[loopbox, fill=green!10] at (12.75,  1.35) (lp1) {
  \textbf{Loop 1: Engineer Coaching}\\[-1pt]
  {\scriptsize retry $+$ context $\to$ cohort $\to$ coaching}
};
\node[loopbox, fill=green!10] at (12.75,  0.0) (lp2) {
  \textbf{Loop 2: Workflow Optimization}\\[-1pt]
  {\scriptsize stage $+$ action $\to$ funnel $\to$ gates}
};
\node[loopbox, fill=green!10] at (12.75, -1.35) (lp3) {
  \textbf{Loop 3: Routing Policy}\\[-1pt]
  {\scriptsize policy $+$ fallback $\to$ A/B $\to$ update}
};

\draw[arr, draw=blue!35]
  (lid.east) to[out=0,in=180] (lp1.west);
\draw[arr, draw=blue!35]
  (ltask.east) to[out=0,in=180] (lp2.west);
\draw[arr, draw=blue!35]
  (lout.east) to[out=0,in=180] (lp3.west);

\end{tikzpicture}
\end{adjustbox}
\caption{The AgentOps feedback architecture.
  \emph{Left}: artifact datasets capture only what agents produce,
  excluding the runtime behavior that determines cost and reliability.
  \emph{Center}: the runtime transaction schema organizes that
  behavior into five layers, namely Identity \& Session, Task Context,
  Routing, Execution, and Outcome Link, assembled locally from
  gateway and transcript sources.
  \emph{Right}: each schema layer enables a concrete feedback
  loop: coaching, workflow optimization, or routing-policy iteration.}
\label{fig:overview}
\end{figure*}

Figure~\ref{fig:overview} summarizes our argument. We propose a
five-stage AgentOps loop (\S\ref{sec:loop}), specify the logical
transaction schema that instantiates it (\S\ref{sec:transactions}),
define three concrete feedback loops with their data requirements
(\S\ref{sec:loops}), report design-probe evidence from \atm
(\S\ref{sec:atm}), and pose three open research problems
(\S\ref{sec:agenda}).

\section{Related Work}
\label{sec:related}

\paragraph{Artifact-centered Agentic SE.}
The dominant paradigm evaluates coding agents through what they
produce: resolve rate on SWE-bench~\cite{swebench2024}, PR
acceptance~\cite{nakashima2026rejected}, and commit
quality~\cite{li2025se3}, across systems such as
SWE-agent~\cite{sweagent2024}, Agentless~\cite{agentless2024}, and
OpenHands~\cite{openhands2025}. This measures capability on defined
tasks but structurally excludes executions that generate cost
without yielding a commit: artifact datasets encode survivorship
bias by construction~\cite{silentci2025, abandonedprs2022}.
AgentLens identifies ``lucky passes'' invisible to outcome-only
evaluation~\cite{agentlens2026}, and failure-trajectory
analysis~\cite{majgaonkar2025trajectories} shows artifact-free
sessions are longer and higher-variance; neither provides
infrastructure to act on these observations.

\paragraph{AgentOps and observability.}
Cloud platforms such as LangSmith, Langfuse,
Helicone~\cite{langfuse, helicone, langsmith}, Arize
Phoenix~\cite{phoenix}, and OpenTelemetry GenAI
instrumentation~\cite{openllmetry, otel-genai} provide prompt and
span tracing; multi-provider gateways such as LiteLLM and
Portkey~\cite{litellm, portkey} add routing and cost logging; a
dedicated AgentOps layer~\cite{agentops2024} instruments agent
sessions end-to-end. These systems assume prompt content can be
forwarded to a centralized collector, an assumption that fails for
coding agents whose payloads contain private source trees,
credentials, and unreleased features~\cite{infraagents2025}.
The present work differs in three specific ways. First, the unit of
analysis is the \emph{task}, not the span or request: outcome linkage
and cost attribution require a task-scoped record that crosses
multiple LLM calls. Second, cross-tool session identity is absent
from OpenTelemetry GenAI spans and Langfuse traces by default, so
cohort analysis across agents (for example, Claude Code and Cursor
used by the same developer) is not supported without custom
instrumentation. Third, the local-first reconciliation model avoids
forwarding source-code payloads to external collectors entirely,
rather than relying on redaction filters.
In their default configurations, these tracing systems record what
happened but do not link execution to downstream outcomes, attribute
cost to tasks, or measure whether an intervention improved
development; such linkage can be added through custom spans and
metadata, but no shared coding-agent transaction substrate exists
that does so out of the box.
Privacy-preserving telemetry~\cite{privacytelemetry2024, mayfly2024}
shows local aggregation can satisfy both governance and analytics,
motivating our local-first approach.

\paragraph{Behavioral analytics as a model.}
Controlled-experiment infrastructure at
scale~\cite{tang2010overlapping, kohavi2013online,
kohavi2012trustworthy} and data-driven metric
development~\cite{deng2016data} show that observational data becomes
actionable only when connected to intervention and
re-measurement, a methodology agentic SE lacks, and for which
runtime transactions are the precondition.

\section{The Missing Feedback Loop}
\label{sec:loop}

A functional feedback loop for AI coding agents requires five stages
that together close the cycle from execution to measurable
improvement. Each stage is necessary: removing any one of them
degrades the system to either a log sink or a dashboard, neither of
which supports deliberate optimization. The runtime layer, comprising
gateway logs and transcript evidence, supplies what artifact-level sources
were not designed to record: failed repairs, abandoned tasks,
fallback chains, and retry-heavy executions that never produce a
commit.

The distinction matters because artifact datasets do not merely
undersample the data they need; they structurally cannot collect it.
They are built from what agents leave behind: commits, pull requests,
passing tests, and benchmark solutions. Sessions that terminate
without producing an artifact leave nothing to scrape. The result is
not a noisy measurement of agent work, but a biased sample of
surviving work. A feedback loop must therefore start at runtime,
where both successful and unsuccessful trajectories are still visible.

\begin{enumerate}
  \item \textbf{Agent execution.} Coding agents (Claude Code, Codex,
    Gemini CLI, OpenCode) run on developer machines, producing gateway
    traffic and local transcripts.
  \item \textbf{Runtime transactions.} A local-first layer reconciles
    both sources into per-request records linked to session and task
    context.
  \item \textbf{Behavioral analytics.} Records are aggregated into
    funnels, cohorts, policy comparisons, and cost attribution.
  \item \textbf{Intervention.} Diagnosis drives coaching, workflow
    templates, routing-policy changes, or guardrails.
  \item \textbf{Measured improvement.} Post-intervention transactions
    are compared against a baseline to verify the change worked.
\end{enumerate}

A system that collects transactions but provides no analytics is a
log sink. A system that diagnoses problems but tracks no outcomes is
a dashboard. The complete five-stage cycle is what turns runtime
observability into an improvement loop.

\section{Runtime Transactions}
\label{sec:transactions}

\paragraph{Definitions.}
We use a small set of terms throughout the schema. A \emph{gateway}
is the local or hosted proxy through which an agent's LLM API calls
pass, observing requests at the network level. A \emph{transcript} is
the agent's on-disk session log, such as the JSONL files Claude Code
writes under \texttt{\textasciitilde/.claude/}, reflecting the session
from the agent's perspective. A \emph{session} is one contiguous agent
invocation; a \emph{task} is the developer-intent unit, such as fixing a bug
or adding an endpoint, and may span multiple sessions. \emph{Routing} is
the gateway's policy-driven choice of provider and model, including
fallbacks; \emph{identity} denotes the pseudonymized developer and
agent tool.

\begin{table*}[t]
\caption{The runtime transaction schema (logical level).
  R~=~required, O~=~optional, R$^{*}$~=~required at task close;
  \full~recorded by default, \partl~partially recoverable,
  \none~absent. \texttt{outcome\_type} $\in$
  \{\texttt{merged\_pr}, \texttt{committed\_patch},
  \texttt{passing\_unreviewed}, \texttt{partial\_artifact},
  \texttt{multi\_session}, \texttt{abandoned}\}.}
\label{tab:schema}
\scriptsize
\setlength{\tabcolsep}{3pt}
\renewcommand{\arraystretch}{0.82}
\begin{adjustbox}{width=0.88\textwidth,center}
\begin{tabular}{@{}llllccl@{}}
\toprule
\textbf{Layer} & \textbf{Field} & \textbf{Type} & \textbf{Req.}
  & \textbf{Gateway} & \textbf{Transcript} & \textbf{Example} \\
\midrule
Identity \& session
  & \texttt{session\_id}      & string        & R & \none  & \full  & \texttt{cc-2026-05-12-a41f} \\
  & \texttt{developer\_id}    & hashed string & O & \partl & \partl & \texttt{a3f2b9c1d8e0} \\
  & \texttt{agent\_tool}      & enum          & R & \partl & \full  & \texttt{claude\_code} \\
\midrule
Task context
  & \texttt{task\_id}         & string        & R & \none  & \none  & \texttt{TASK-1287} \\
  & \texttt{task\_type}       & enum          & O & \none  & \partl & \texttt{bugfix} \\
  & \texttt{project}          & string        & O & \none  & \full  & \texttt{payments-api} \\
\midrule
Routing
  & \texttt{auth\_mode}       & enum          & R & \partl & \partl & \texttt{subscription} \\
  & \texttt{policy\_id}       & string        & O & \full  & \none  & \texttt{cost-aware-v3} \\
  & \texttt{requested\_model} & string        & R & \full  & \full  & \texttt{sonnet-primary} \\
  & \texttt{resolved\_model}  & string        & R & \full  & \none  & \texttt{provider-b/32b} \\
  & \texttt{fallback\_chain}  & list[string]  & O & \full  & \none  & \texttt{[primary, fb-1]} \\
\midrule
Execution
  & \texttt{request\_ts}      & timestamp     & R & \full  & \full  & \texttt{2026-05-12T09:14:03Z} \\
  & \texttt{in/out\_tokens}   & int           & R & \full  & \full  & \texttt{9812/1204} \\
  & \texttt{status}           & enum          & R & \full  & \none  & \texttt{rate\_limited} \\
  & \texttt{retry\_index}     & int           & O & \full  & \partl & \texttt{2} \\
  & \texttt{stage / action\_type} & enum      & O & \none  & \partl & \texttt{repair/edit} \\
  & \texttt{cost\_usd}        & decimal       & O & \full  & \none  & \texttt{0.0421} \\
\midrule
Outcome
  & \texttt{outcome\_link}    & artifact ref  & O & \none  & \none  & \texttt{repo@9f3e2c1} \\
  & \texttt{outcome\_type}    & enum          & R$^{*}$ & \none & \none & \texttt{merged\_pr} \\
\bottomrule
\end{tabular}
\end{adjustbox}
\end{table*}

Token counts answer \emph{how much} but leave \emph{why},
\emph{what happened}, and \emph{whether the routing decision was
correct} unrecorded. A request consuming 10,000 input tokens might
be a successful refactor, a failed retry loop, or a policy-blocked
attempt; the count is identical in all three, which is why token
accounting alone cannot support any feedback loop.

The observability challenge is one of \emph{structural absence}:
artifact datasets capture only the tail of a session: the commit,
the PR, the passing test. The schema in Table~\ref{tab:schema}
organizes the missing context into five layers, from the innermost
record outward: \textbf{Execution} (per-request timing, tokens, and
status), \textbf{Routing} (policy, requested vs.\ resolved model,
fallback chain), \textbf{Task Context} (task identifier, type, and
project), \textbf{Identity \& Session} (cross-tool session id and
pseudonymized developer), and \textbf{Outcome Link} (the artifact
produced, if any). Each layer restores context that artifacts
structurally exclude; together they form the substrate required by
all three feedback loops in \S\ref{sec:loops}.

Gateway logs supply routing truth but no session or task context;
transcripts supply session context but no routing truth;
reconciliation is what creates analytical value. The Outcome layer is
the most consequential gap: neither source records whether a session
produced an accepted artifact, so no loop can compute cost per
accepted patch or completion rate. Identity gaps block Loop~1,
task-context gaps block Loop~2, and the absent outcome layer removes
the denominator for all three.

\paragraph{Mapping to existing tools.}
The Execution layer maps directly onto existing telemetry:
\texttt{request\_ts}, token counts, and model fields correspond to
OpenTelemetry GenAI span attributes~\cite{otel-genai} and Langfuse
observation fields~\cite{langfuse}, while \texttt{policy\_id},
\texttt{resolved\_model}, \texttt{fallback\_chain}, and
\texttt{cost\_usd} correspond to LiteLLM and Helicone gateway
metadata~\cite{litellm, helicone}. The remaining fields, namely
cross-tool \texttt{session\_id}, \texttt{task\_id}, \texttt{task\_type},
\texttt{stage}, and the Outcome layer, have no default counterpart
in any system we examined; making them first-class is the substance
of this proposal.

\section{Three Feedback Loops}
\label{sec:loops}

The three subsections below specify what each loop can accomplish and
identify the infrastructure gap that currently prevents it. The point
is not that these analytics are novel in isolation, but that agentic
software engineering lacks the transaction layer needed to apply them
to coding-agent behavior.

\subsection{Loop 1: Engineer Coaching}

Runtime traces expose behavioral fingerprints invisible to
outcome-only analysis: broad context selection, absent plan steps,
retries that add no new information, and high abort rates on task
types where peers succeed~\cite{agentlens2026}. We observe these
patterns in prototype deployment and hypothesize that cohort analysis
can link them to poor outcomes and drive targeted interventions:
plan-first coaching, scoped-context guidelines, or task-decomposition
templates. Two engineers producing equal numbers of accepted patches
are distinguishable only at the runtime level when one consistently
consumes five times the tokens getting there. The infrastructure gap
is identity: a cross-tool session identifier, task type, and
acceptance signal are recorded by default by none of the systems
examined.

\subsection{Loop 2: Workflow Optimization}

With stage boundaries and action sequences instrumented, funnel
analysis can compute per-stage conversion, sequence mining can
identify failure paths such as
``edit $\to$ skip test $\to$ repair $\to$ abort,'' and survival
analysis can locate the stage most associated with
abandonment~\cite{majgaonkar2025trajectories, lu2025autonomous}.
Interventions follow directly: scoped context packs, forced test
gates, or repair-loop limits. The gap is granularity: these signals
are not exposed by default; custom spans add them only as per-team
effort, with no portable dataset. As a result, teams can inspect an
individual failed transcript, but cannot ask which workflow stage most
often predicts abandonment across a cohort.

\subsection{Loop 3: Routing Policy Iteration}

With per-request outcomes linked to task completion, policy cohort
analysis can compare total cost per completed task, fallback rate,
and success rate across policy groups, with A/B assignment for
controlled comparison~\cite{kohavi2012trustworthy, deng2016data,
frugalgpt2023, routellm2024}. This loop is closest to achievable
today, since gateways already record most routing fields. The gap is
the outcome join: cost per completed task requires
\texttt{outcome\_link}, absent from every source we examined. Without
it, a policy that cuts per-request price while increasing retries
looks like a gateway saving but a task-level cost increase, the
failure mode of Vignette~2.

\section{\atm as a Design Probe}
\label{sec:atm}

We developed Agent Token Manager (\atm) to evaluate the schema. \atm
runs as a local background process: it intercepts gateway traffic in
key-mode and reads transcript files from disk, merging both into a
unified session record without forwarding prompt content externally.
It is a design probe, not a validation study: the observations below
are existence evidence from one small deployment, not population
estimates, and do not yet show that any feedback loop improves
outcomes. Its purpose is narrower: to test whether the missing fields
in Table~\ref{tab:schema} correspond to real blind spots in deployed
agent use.

\paragraph{Methodology.}
\atm identifies sessions by reading agent-specific transcript
directories (e.g., \texttt{\textasciitilde/.claude/},
\texttt{\textasciitilde/.codex/}); each entry carries a timestamp,
token counts, and a session boundary marker. Task types were assigned
by mapping project directories to three manual labels (systems,
ml-research, writing). Gateway requests were captured only for
key-mode authentication via a local proxy. Key-mode reconciliation
joins a gateway record and a transcript entry when their timestamps
fall within a small window and their model names and token counts
agree; subscription-mode requests are classified as transcript-only
because no gateway record exists by construction. We have not measured
this heuristic's error rate. Reporting false-match, missed-match, and
ambiguous-match rates against a labeled sample is a prerequisite for
trusting any feedback loop built on reconciled records, and is a
concrete next step for validation.

\paragraph{Deployment scale.}
Over a 93-day deployment, \atm instrumented four agents, namely
Claude Code, Codex, Cursor, and OpenClaw, spanning 23 projects and three
task-type categories: 51 sessions, 6{,}775 transcript-side requests,
and \$552 in agent spend, exposing two structural gaps that motivate
the schema directly.

\paragraph{The subscription blind spot.}
The gateway captured only 91 key-mode requests; the remaining
6{,}684 ran under subscription-mode authentication, which produces
no gateway record at all. From a gateway-only view, 98.7\% of all
requests in this deployment are invisible. Subscription mode is a
common default for Claude Code and Codex, so much coding-agent traffic
structurally evades gateway accounting.

\paragraph{Model-substitution divergence.}
In a controlled 60-request key-mode trace, the gateway attributed all
requests to MiniMax-M2.7, the provider selected by the routing policy,
while the Claude Code transcript recorded the same requests as
\texttt{claude-sonnet-4-6}, the model the agent believed it was
addressing. Neither source alone recovers the correct picture:
gateway truth (routing, resolved provider, and cost) and transcript
truth (agent-perceived model and session context) diverge. A
single-source system silently adopts one and discards the other.
Reconciling both is required before any routing-policy analysis is
valid.

\paragraph{Session behavioral distribution.}
Session turn counts ranged from 1 to 526 (mean 107; 71\% exceeded
15 turns), far above what benchmark task runners produce, reflecting
the multi-step repair, context-gathering, and recovery sequences of
real developer sessions, precisely the states the \texttt{stage} and
\texttt{action\_type} fields capture.

Together these confirm that no single source closes the three loops:
the subscription blind spot blocks Loop~3, and substitution
divergence invalidates routing attribution from gateway logs alone.
They also show why the schema is task-scoped rather than merely
request-scoped: the operational question is not which request was
cheap, but whether the complete task became cheaper, faster, and more
likely to complete.

\section{Open Research Problems}
\label{sec:agenda}

Three problems must be solved before these loops close at scale.

\textbf{Trace reconciliation without shared request identifiers.}
Probabilistic record linkage on timestamps, model names, and token
counts is theoretically grounded~\cite{fellegi1969theory,
splink2022}, and execution log mining~\cite{lou2010mining,
nandi2016anomaly} and entity resolution~\cite{bhattacharya2006query,
bellare2012active} supply algorithmic foundations. The agent setting
adds a challenge they lack: a transcript entry may have no gateway
record \emph{because} the agent ran in subscription mode, so absence
carries diagnostic meaning rather than signaling a failed join.
Distinguishing the two cases and quantifying false-, missed-, and
ambiguous-match rates remains open.

\textbf{Outcome linking across asynchronous systems.}
A session may produce a commit hours after its requests complete, or
no artifact at all; linking transactions to downstream tests,
reviews, and acceptance decisions without a shared key is an open
record-linkage problem~\cite{linktransformer2024}.
Each \texttt{outcome\_type} in Table~\ref{tab:schema} differs:
\texttt{committed\_patch} matches by repository, branch, files
touched, and temporal proximity; \texttt{merged\_pr} additionally
requires following review to a merge or rejection;
\texttt{passing\_unreviewed} and \texttt{partial\_artifact} capture
useful intermediate results that fall short of a merge;
\texttt{multi\_session} requires a stable \texttt{task\_id} across
invocations; and \texttt{abandoned} is hardest: indistinguishable at
runtime from a session whose artifact has not yet appeared, forcing
timeout heuristics or explicit agent-side signaling.

\textbf{Intervention causality.}
Whether a coaching session or policy update caused genuine
improvement is confounded by task-mix shifts, model updates, and
learning curves. Causal designs for agentic
SE, such as difference-in-differences and synthetic control, are largely
unstudied, since the treatment unit is a developer or routing policy,
not a randomized user; without them, teams cannot separate real
effects from favorable workload shifts.

\section{Limitations and Governance}
\label{sec:limits}

\paragraph{Privacy and developer surveillance.}
Local transcripts contain source code, file paths, and prompts that
may include credentials or proprietary context. Even without
external forwarding, local aggregation of turn counts, retry rates,
and per-developer cost creates productivity signals open to misuse
for ranking or surveillance.
We propose these as deployment defaults: opt-in collection with
per-developer revocation; a minimum cohort size so no reported
aggregate resolves to an individual; separation of developer-facing
coaching from aggregate-only management reporting; retention limits on
raw transcripts; and an explicit prohibition on using these metrics
for individual performance ranking.

\paragraph{Instrumentation burden and generalizability.}
The probe required agent-specific parsing per transcript format, and
task-type labels were manually assigned from project
directories, too coarse for generalizable analytics; automatic
inference is future work. The small single-group deployment may not
generalize to enterprise teams, on-call workflows, or CI/CD agents
lacking local transcripts. Generality requires deployments across
teams and organizations with differing codebases, agent mixes, and
routing policies; we are preparing the schema and \atm prototype for
release, and view a multi-team deployment with at least one
closed-loop intervention, namely measuring cost per completed task
before and after a routing-policy update, as the natural next study.

\paragraph{Outcome ambiguity.}
We define success conservatively as a linked accepted artifact
(merged PR or committed patch), but many outcomes are partial, which
the \texttt{outcome\_type} values of Table~\ref{tab:schema} capture:
\texttt{passing\_unreviewed} (passes tests, awaits review),
\texttt{partial\_artifact} (a useful intermediate result, no commit),
\texttt{multi\_session} (succeeds only across sessions), and
\texttt{abandoned} (assigned after a linkage timeout).

\section{Conclusion}
\label{sec:conclusion}

Artifact datasets and token accounting capture only what coding
agents produce, not why sessions fail, where workflows break down, or
whether routing decisions reduced total task cost. The runtime
behaviors that most influence cost, reliability, and productivity
are precisely the ones excluded by default: failed retries, fallback
chains, abandoned sessions, and multi-step repair loops that never
become commits.

We have argued that a runtime transaction schema, assembled locally
from gateway and transcript sources, is the necessary precondition
for the three feedback loops agentic SE needs: engineer coaching,
workflow optimization, and routing-policy iteration. Our design
probe, \atm, shows that task-scoped records are feasible under a
local-first privacy model and surfaces the open problems that must be
solved before these loops can operate at scale: reconciliation
accuracy, outcome linking, and intervention causality.

Our central position is this: runtime transaction infrastructure
should become a core evaluation and operations layer for agentic
software engineering, not a post-hoc add-on. Without it, engineering
teams are optimizing systems they cannot observe, and the field is
evaluating agents on a biased sample of the sessions that matter.

\bibliographystyle{ACM-Reference-Format}
\bibliography{refs}

\end{document}